# CLP approaches to 2D angle placements

Tomasz Szczygieł

Politechnika Śląska, Instytut Automatyki, Akademicka 16, 44-100 Gliwice, Poland
tszczygiel@inetia.pl

**Abstract.** The paper presents two CLP approaches to 2D angle placements, implemented in CHIP v.5.3. The first is based on the classical (rectangular) cumulative global constraint, the second on the new trapezoidal cumulative global constraint. Both approaches are applied to a specific presented.

## 1  Introduction

The problem discussed has its roots in the packing and transportation of high-current enclosed conductors and high-current enclosed bus bars in the most economical way. The producer of those elements is interested in packing the largest number of elements delivered to the customer in the space available. Sometimes the cost of transporting those elements exceed the cost of producing them. The packing of those elements into containers may be modeled as a three-dimensional angle packing problem; however the problem of packing them into long-load trailer may best be modeled as two-dimensional angle packing problem. This paper present four CHIP solutions for the two-dimensional angle packing problem: the angles may be packed with or without rotation, the packing may be done using either the rectangular or the trapezoidal cumulative global constraint.

## 2  Angle packing with no rotation

The angles to be packed into a larger square or rectangle have fixed orientations. An earlier paper [6] discussed the details. This angle packing problem may be formulated as a puzzle problem or as a bin packing problem (see [8]). For the puzzle problem, a CLP solution does not include a meta predicate for optimization of some cost. However for the bin packing problem, the CLP solution must rely upon predicates like *min_max*. Because the last problem is more general, it will be discussed in detail.

**Example I:** 10 small angles are to be packed into a large rectangle so that none of them is overlapping any other. Table 1 gives the data for the problem.

**Table 1.** Data for angle packing problem with no rotation

| No | List of angle sizes | No | List of angle sizes |
|----|---------------------|----|---------------------|
| 1  | [2,4,3,1]           | 6  | [1,2,5,5]           |
| 2  | [2,2,1,3]           | 7  | [6,2,2,3]           |
| 3  | [1,3,3,2]           | 8  | [4,2,2,1]           |
| 4  | [2,1,4,3]           | 9  | [3,1,1,4]           |
| 5  | [1,7,2,2]           | 10 | [3,2,1,1]           |

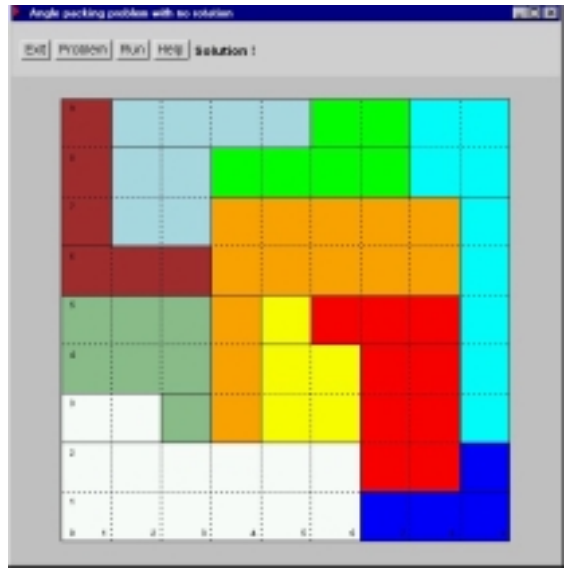

**Fig.1**. A solution for the angle packing problem with no rotation.

### 2.1 Classical cumulative approach

The idea of solving the angle packing problem by the classical cumulative approach is based on solution of the rectangle packing problem with some additional constrains (more details - see [6]). At the beginning the angles are divided into rectangles. This makes predicate *gen_rect/4* as follows:

```
gen_rect([],[],[],[]).
gen_rect([DL|DT],[DXH1,DXH2|DXT],[DYH1,DYH2|DYT],[LLH|LLT]):-
         position(DL,DXH1,DXH2,DYH1,DYH2,LLH),
         gen_rect(DT,DXT,DYT,LLT).
```

Predicate *position/6* transforms the list of angle sizes DL in the following way:
- it divides each angle into two rectangles and describes the sizes of the rectangles
- it fixed the angle orientation.

```
position([A,B,C,D],DXH1,DXH2,DYH1,DYH2,1):-
        B #> D,A #< C,DXH1 is C, DXH2 is A,
        DYH1 is D,DYH2 is B-D.
```

The discussed additional constrains merge two rectangles into an angle. These constrains are generated automatically by the predicate *constr_rect/5*. The predicate is defined as follows:

```
constr_rect([],[],[],[],[]).
constr_rect([LXH1,LXH2|LXT],[LYH1,LYH2|LYT],
        [DXH1,DXH2|DXT],[DYH1,DYH2|DYT],[LLH|LLT]):-
        if LLH #= 1 then LXH1+DXH1 #= LXH2+DXH2,
        if LLH #= 1 then LYH1 #= LYH2+DYH2,
        if LLH #= 2 then LXH1 #= LXH2,
        if LLH #= 2 then LYH1 #= LYH2+DYH2,
        constr_rect(LXT,LYT,DXT,DYT,LLT).
```

The final program solving example I is given below:

```
run:-
   EndX :: 1..9,
   EndY :: 1..9,
   High1:: 1..9,
   High2:: 1..9,
   min_max((data(Data),
        gen_rect(Data,DX,DY,LL),
        gen_lists(DX,DY,LX,LY,DF),
        constr_rect(LX,LY,DX,DY,LL),
        diffn(DF),
        cumulative(LX,DX,DY,unused,unused,High1,EndX,unused),
        cumulative(LY,DY,DX,unused,unused,High2,EndY,unused),
        append(LX,LY,XY),
        labeling(XY)),EndX+EndY).
```

## 2.2 Cumulative trapeze approach

The idea of solving the angle packing problem by cumulative trapeze approach is based on the dividing angles into rectangles too. But in this approach additional constrains are not needed, because *cumulative_trapeze* is a standard predicate embedded in CHIP and merging two rectangles into an angle is done automatically. Now the program solving the same example I is as follows:

```
run:-
    EndX :: 1..9,
    EndY :: 1..9,
    min_max((data(Data),
            gen_lists(Data,1,LX,LY,EX,EY,TX,TY,TRX,TRY,DF),
            diffn(DF),
            cumulative_trapeze(TX,TRX,EndX),
            cumulative_trapeze(TY,TRY,EndY),
            append(LX,LY,XY),
            labeling(XY)),EndX+EndY).
```

Next, the angles are divided into rectangles by the predicate *gen_lists/11*. This predicate is generating terms for the global constrain *cumulative_trapeze*. The predicate is defined as follows:

```
gen_lists([],_,[],[],[],[],[],[],[],[],[]).
gen_lists([DH|DT],N,[LXH|LXT],[LYH|LYT],
        [EXH|EXT],[EYH|EYT],[TXH|TXT],[TYH|TYT],
        [TRXH1,TRXH2|TRXT],[TRYH1,TRYH2|TRYT],
        [DFH1,DFH2|DFT]):-
    position(DH,DXH,DYH,SXH1,SXH2,SYH1,SYH2,
             DX1,DX2,DY1,DY2,EXH1,EXH2,EYH1,EYH2),
    LXH    :: 0..9,
    LYH    :: 0..9,
    EXH    :: 1..9,
    EYH    :: 1..9,
    [G1,G2,G3,G4] :: 0..9,
    [F1,F2,F3,F4] :: 0..9,
    TXH = task(N,LXH,DXH,EXH),
    TYH = task(N,LYH,DYH,EYH),
    TRXH1 = trap(N,1,SXH1,DX1,EXH1),
    TRXH2 = trap(N,2,SXH2,DX2,EXH2),
    TRYH1 = trap(N,1,SYH1,DY1,EYH1),
    TRYH2 = trap(N,2,SYH2,DY2,EYH2),
    DFH1  = [G1,G2,G3,G4],
    DFH2  = [F1,F2,F3,F4],
    list_diff(DH,LXH,LYH,DFH1,DFH2),
    N1 is N+1,
    gen_lists(DT,N1,LXT,LYT,EXT,EYT,TXT,TYT,TRXT,TRYT,DFT).
```

The predicate *position/15* is – as a result of analyzing the list of angle sizes - assigning values of variables in the global constrain *cumulative_trapeze*.

```
position([A,B,C,D],DXH,DYH,SXH1,SXH2,SYH1,SYH2,DX1,DX2,
        DY1,DY2,EXH1,EXH2,EYH1,EYH2):-
        B #< D, A #< C, DXH is C, DYH is D, SXH1 is D,
        SXH2 is B, DX1 is A, DX2 is C-A, EXH1 is D,
        EXH2 is B, SYH1 is A, SYH2 is C, DY1 is D-B,
        DY2 is B, EYH1 is A, EYH2 is C.
```

The predicate *list_diff/5* prepare list of variables for standard predicate *diffn*.

```
list_diff([A,B,C,D],LXH,LYH,[G1,G2,G3,G4],[F1,F2,F3,F4]):-
           A #< C, B #> D, G1 is LXH, G2 is LYH+B-D,
           G3 is C-A, G4 is D, F1 is LXH+C-A,
           F2 is LYH, F3 is A, F4 is B.
```

To keep this paper short , only one definition of predicate *position/15* and *list_diff/5* are presented.

## 3 Angle packing with rotation n*90°

Now, the rectangles may be rotated before placement. The problem is considered hard since the angle orientation are initially not fixed. In this section problem may be formulated similarly way as a puzzle problem or as a bin packing problem. The solutions of this problem has been supported by detailed study of Examples II.

**Example II:** 4 small angles are to be packed with rotation n*90° and mirror reflection into a large rectangle so that none of them is overlapping any other. Table 2 gives the data for the problem.

**Table 2.** Data for angle packing problem with no rotation

| No | List of angle sizes | No | List of angle sizes |
|---|---|---|---|
| 1 | [3,7,7,2] | 3 | [2,5,4,3] |
| 2 | [2,10,3,7] | 4 | [3,8,5,2] |

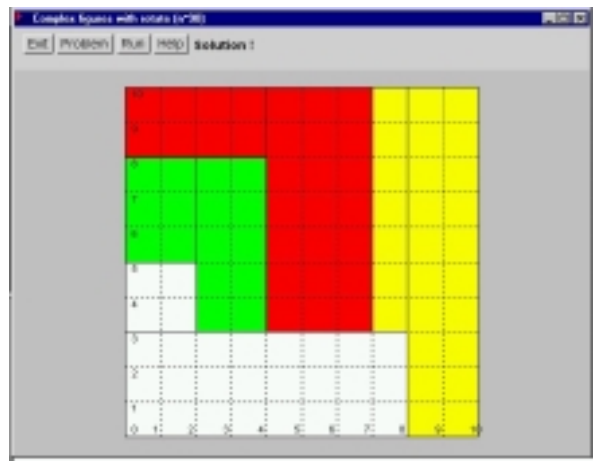

**Fig.2**. A solution for the angle packing problem with rotation.

### 3.1 Classical cumulative approach

The idea of solving the angle packing problem with rotation by the classical cumulative approach is based on the fact that each angle is included in a rectangle. At the beginning the angle are divided into two rectangles (more details - see [6]). This makes predicate *gen_rect/3* as follows:

```
gen_rect([],[],[]).
gen_rect([D0L|D0T],[DH1,DH2|DT],[DX,DY|DXY]):-
        div_angle(D0L,DH1,DH2,DX,DY),
        gen_rect(D0T,DT,DXY).
div_angle([A,B,C,D],DH1,DH2,DX,DY):-
```

```
        DX  is C,
        DY  is B,
        E   is B-D,
        DH1 = C*D,
        DH2 = A*E.
```

Then those rectangles are placed with rotation on the plane; and additional constraint merge two rectangles into an angle. This is done by the predicates *constrain_rect/5* and *gen_lists/6* as follows.

```
constrain_rect([],[],[],[],[]).
constrain_rect([LXH1,LXH2|LXT],[LYH1,LYH2|LYT],
               [DXH1,DXH2|DXT],[DYH1,DYH2|DYT],[DX,DY|DT]):-
        STX :: 0..9,
        STY :: 0..9,
        ENX :: 1..10,
        ENY :: 1..10,
        dll_y(LYH1,LYH2,DYH1,DYH2,STY,ENY),
        dll_x(LXH1,LXH2,DXH1,DXH2,STX,ENX),
        [A,B] :: [DX,DY],
        (DX \= DY -> A #\= B; true),
        ENX #= A+STX,
        ENY #= B+STY,
        constrain_rect(LXT,LYT,DXT,DYT,DT).

dll_x(LXH1,LXH2,DXH1,DXH2,STX,ENX):-
        LXH1 #<= LXH2,
        LXH1+DXH1 #>= LXH2+DXH2,
        STX #= LXH1,
        ENX #= LXH1+DXH1.
dll_y(LYH1,LYH2,DYH1,DYH2,STY,ENY):-
        LYH1 #<= LYH2,
        LYH1+DYH1 #>= LYH2+DYH2,
        STY #= LYH1,
        ENY #= LYH1+DYH1.
```

Only one definition of predicate *dll_x/6* and *dll_y/6* are presented.

```
gen_lists([],[],[],[],[],[]).
gen_lists([X1*Y1,X2*Y2|T],[LXH1,LXH2|LXT],[LYH1,LYH2|LYT],
        [DXH1,DXH2|DXT],[DYH1,DYH2|DYT],[DFH1,DFH2|DFT]):-
        LXH1 :: 0..9,
        LYH1 :: 0..9,
        LXH2 :: 0..9,
        LYH2 :: 0..9,
        [DXH1,DYH1] :: [X1,Y1],
        [DXH2,DYH2] :: [X2,Y2],
        (X1 \= Y1 -> DXH1 #\= DYH1; true),
        (X2 \= Y2 -> DXH2 #\= DYH2; true),
        (DXH1 = X1 -> DXH2 #= X2; DXH2 #= Y2),
        (DYH1 = Y1 -> DYH2 #= Y2; DYH2 #= X2),
         append([LXH1,LYH1],[DXH1,DYH1],DFH1),
         append([LXH2,LYH2],[DXH2,DYH2],DFH2),
         gen_lists(T,LXT,LYT,DXT,DYT,DFT).
```

Program solving the example II is as follows:

```
run:-
      data(Data0),
      gen_rect(Data0,Data,DXY),
      EndX :: 1..10,
      EndY :: 1..10,
      HighX :: 1..10,
      HighY :: 1..10,
      min_max((gen_lists(Data,LX,LY,DX,DY,DF),
              constrain_rect(LX,LY,DX,DY,DXY),
              diffn(DF,unused,unused,[EndX,EndY]),
          cumulative(LX,DX,DY,unused,unused,HighX,EndX,unused),
          cumulative(LY,DY,DX,unused,unused,HighY,EndY,unused),
              append(LX,LY,LXY),
              append(DX,DY,DDXY),
              append(LXY,DDXY,XY),
              labeling(XY)),EndX+EndY).
```

## 3.2 Cumulative trapeze approach

Now the program solving the same example II is as follows:

```
  run:-
     EndX :: 1..10,
     EndY :: 1..10,
     min_max((data(Data),
             gen_lists(Data,1,LX,LY,TX,TY,TRX,TRY,DF),
             diffn(DF),
             cumulative_trapeze(TX,TRX,EndX),
             cumulative_trapeze(TY,TRY,EndY),
             append(LX,LY,XY),
             labeling(XY)),EndX+EndY).
```

In this point predicate gen_lists/9 simultaneously:
- divides the angles into the rectangles
- generates terms for the global constrain *cumulative_trapeze*.
- merges two rectangles into an angle
- generates list for global constrain d*iffn*.
The predicate is defined as follows:

```
  gen_lists([],_,[],[],[],[],[],[],[]).
  gen_lists([[A,B,C,D]|DT],N,
       [LXH|LXT],[LYH|LYT],[TXH|TXT],[TYH|TYT],
       [TRXH1,TRXH2|TRXT],[TRYH1,TRYH2|TRYT],
       [[LX1,LY1,DDX1,DDY1],[LX2,LY2,DDX2,DDY2]|DFT]):-
     F is B-D,
     E is C-A,
     LXH :: 0..9, LYH :: 0..9, DXH :: 1..10, DYH :: 1..10,
     EXH :: 1..10, EYH :: 1..10,
     SXH1 :: 1..10, DX1 :: 1..10, EXH1 :: 1..10,
     SXH2 :: 1..10, DX2 :: 1..10, EXH2 :: 1..10,
```

```
        SYH1 :: 1..10, DY1 :: 1..10, EYH1  :: 1..10,
        SYH2 :: 1..10, DY2 :: 1..10, EYH2  :: 1..10,
        DDX1 :: 1..10, DDY1 :: 1..10,
        DDX2 :: 1..10, DDY2 :: 1..10,
        LX1 :: 0..9, LY1 :: 0..9,
        LX2 :: 0..9, LY2 :: 0..9,
        pozition(A,B,C,D,E,F,LXH,DXH,EXH,LYH,DYH,EYH,
                 SXH1,DX1,EXH1,SXH2,DX2,EXH2,
                 SYH1,DY1,EYH1,SYH2,DY2,EYH2,
                 LX1,LY1,DDX1,DDY1,LX2,LY2,DDX2,DDY2),
    TXH = task(N,LXH,DXH,EXH),
    TYH = task(N,LYH,DYH,EYH),
    TRXH1 = trap(N,1,SXH1,DX1,EXH1),
    TRXH2 = trap(N,2,SXH2,DX2,EXH2),
    TRYH1 = trap(N,1,SYH1,DY1,EYH1),
    TRYH2 = trap(N,2,SYH2,DY2,EYH2),
    N1 is N+1,
    gen_lists(DT,N1,LXT,LYT,TXT,TYT,TRXT,TRYT,DFT).
```

The predicate *pozition/32* is a definition of all angle orientation on the plane and determines all variables in the global constrain *cumulative_trapeze* and *diffn* by the coordinates of a chosen point of the rectangle including angles. The predicate is given below:

```
pozition(A,B,C,D,E,F,LXH,DXH,EXH,LYH,DYH,EYH,
         SXH1,DX1,EXH1,SXH2,DX2,EXH2,SYH1,DY1,EYH1,
         SYH2,DY2,EYH2,LX1,LY1,DDX1,DDY1,
         LX2,LY2,DDX2,DDY2):-
    DXH is C, EXH is LXH+C, DYH is B, EYH is LYH+B,
    SXH1 is D, DX1 is E, EXH1 is D,
    SXH2 is B, DX2 is A, EXH2 is B,
    SYH1 is A, DY1 is F, EYH1 is A,
    SYH2 is C, DY2 is D, EYH2 is C,
    LX1 is LXH, LY1 is LYH+F, DDX1 is E, DDY1 is D,
    LX2 is LXH+E, LY2 is LYH, DDX2 is A, DDY2 is B.
```

To keep this presentation short, only one definition of the predicate position/32 is presented. Other cases may be formulated in a similar way.

## 4 Conclusions

Table 3 shows the time needed for finding the solution for the ten angle packing problem with no rotation. In both cases programs using the classical cumulative constraint and additional *constr_rect/5* predicate need less time to get the solution than programs with only the cumulative trapeze constraint.

**Table 3.** Times for 10 angle with no rotation

| Without optimization | | With optimization | |
|---|---|---|---|
| Cumulative [hh:mm:ss.ms] | Cumultive_trapeze [hh:mm:ss.ms] | Cumulative [hh:mm:ss.ms] | Cumulative_trapeze [hh:mm:ss.ms] |
| 00:00:02.033 | 00:00:04.306 | 00:00:02.083 | 00:00:05.208 |

The Table 4 shows the time needed for finding the solution for various angle packing problem with rotation and without optimization. For the program using the classical cumulative constraint less time is needed to get the solution than for the programs with the cumulative trapeze constraint. The program using the classical cumulative exceed 2 hours limit in case with ten angles. The programs using cumulative_trapeze exceed this limit earlier – for nine angles.

**Table 4.** Times for various angles with rotation- without optimization

| Number of angle | Cumulative [hh:mm:ss.ms] | Cumulative_trapeze [hh:mm:ss.ms] |
|---|---|---|
| 4 | 00:00:00.000 | 00:00:00.030 |
| 5 | 00:00:00.130 | 00:00:00.711 |
| 6 | 00:00:00.661 | 00:00:02.724 |
| 7 | 00:00:16.774 | 00:01:28.347 |
| 8 | 00:04:29.407 | 00:27:00.460 |
| 9 | 01:58:33.940 | >02:00:00.000 |
| 10 | >02:00:00.000 | >02:00:00.000 |

The Table 5 shows the time needed for finding the solution for various angle packing problem with rotation and with optimization. Now too, the programs including classical cumulative constraint works quicker than programs including the cumulative trapeze constraint.

**Table 5.** Times for various angles with rotation- with optimization

| Number of angle | Cumulative [hh:mm:ss.ms] | Cumulative_trapeze [hh:mm:ss.ms] |
|---|---|---|
| 4 | 00:00:01.221 | 00:00:08.532 |
| 5 | 00:00:22.973 | 00:00:59.596 |
| 6 | 00:03:29.976 | 00:21:01.414 |
| 7 | 00:13:45.908 | >02:00:00.000 |
| 8 | >02:00:00.000 | >02:00:00.000 |
| 9 | >02:00:00.000 | >02:00:00.000 |
| 10 | >02:00:00.000 | >02:00:00.000 |

Presented results support the thesis: that more simple constraints (like cumulative) work faster, but more sophisticated constraints (like cumulative trapeze) make program writing easier. All computation experiments were based on Pentium II / 300MHz with 128MB RAM station.

## 5 Acknowledgment

The author is grateful to Prof. A. Niederliński for his support, help and encouragement.